\documentclass[3p,times]{elsarticle}
\usepackage{xcolor}
\usepackage{dsfont}
\usepackage{amsmath}
\usepackage{amssymb}
\usepackage{graphicx}
\usepackage{enumerate}
\usepackage{hyperref}
\usepackage[font=small,skip=1.5pt]{caption}
\usepackage{cleveref}
\usepackage{amsthm}
\theoremstyle{definition}
\newtheorem{Theorem}{Theorem}[section]
\newtheorem{Proposition}{Proposition}[section]

\newtheorem{Example}{Example}
\usepackage{algorithm}
\usepackage{epstopdf}
\usepackage{algpseudocode}

\newenvironment{Proof}{\paragraph{Proof:}}{\hfill$\blacksquare$}

\def\arrvline{\hfil\kern\arraycolsep\vline\kern-\arraycolsep\hfilneg}
\journal{}
\begin{document}

\begin{frontmatter}

\title{A computational weighted finite difference method for American and barrier options in subdiffusive Black-Scholes model}
\author[label1]{Grzegorz Krzy\.zanowski}
\ead{grzegorz.krzyzanowski@pwr.edu.pl}

\author[label1]{Marcin Magdziarz}
\ead{marcin.magdziarz@pwr.wroc.pl}

\address{Hugo Steinhaus Center,
Faculty of Pure and Applied Mathematics, Wroclaw University of Science and Technology
50-370 Wroclaw, Poland}

\begin{abstract}
Subdiffusion is a well established phenomenon in physics. In this paper we apply the subdiffusive dynamics to analyze financial markets. We focus on the financial aspect of time fractional diffusion model with moving boundary i.e. American and barrier option pricing in the subdiffusive Black-Scholes (B-S) model. Two computational methods for valuing American options in the considered model are proposed -
the weighted finite difference (FD) and the Longstaff-Schwartz method.
In the article it is also shown how to valuate numerically wide range of barrier options using the FD approach.
\end{abstract}


\end{frontmatter}



\section{Introduction}
Subdiffusion is the standard 
tool in the analysis of complex systems. In physics it is manifested by the celebrated Fractional Fokker-Planck
equation. This equation was derived from the continuous-time random walk scheme
with heavy-tailed waiting times. Equivalently it can be described in terms of subordination, where the standard diffusion process is time-scaled by the
so-called inverse subordinator. Nowadays the subdiffusion is a very well established phenomenon with many real-life examples - e.g. diffusion in percolative
and porous systems, charge carrier transport in amorphous semiconductors, transport on fractal geometries. (see \cite{Gajda, orzel} and references in them)

In this paper we apply the subdiffusive dynamics to analyze financial markets. Option pricing is the core content of modern finance and has fundamental meaning for global economy. By the recent announcement of Futures Industry
Association trading activity in the global exchange-traded derivatives markets, in $2019$ reached a record of $34,47$ billion contracts, where $15,23$ billion of them were options contracts \cite{fia}.
The value of the global derivatives markets is estimated  $700$ trillion dollars to upwards of $1,5$ quadrillion dollars (including so called shadow derivatives) \cite{market}.

American option is one of the most popular financial derivatives (e.g. most listed options in the USA are Americans) \cite{Dodan2}. It is widely accepted by investors for its flexibility of
exercising time (see e.g. \cite{LSswap}). Barrier options are the simplest of all exotic options traded on financial markets \cite{weron}.
This kind of security is a standard vanilla option which begins to be valid if the price of the underlying asset hits predetermined barrier (or barriers) before the maturity.
They have become increasingly popular due to
the lower costs and the ability to match speculating or hedging needs more closely than their vanilla equivalents.
Moreover, barrier options play an important role in managing and modeling risks
in finance as well as in refining insurance products such as variable
annuities and equity-indexed annuities \cite{Bar2, Bar1}.

Over the last two decades the B-S model has been increasingly attracting interest  as  effective tool of the options valuation. The model was of such great importance that
the authors were awarded the Nobel Prize for Economics in 1997. The classical model was generalized in order to weaken its strict assumptions, allowing such features as stochastic interest \cite{Merton}, regime switching  model \cite{com1},
stochastic volatility \cite{Hull},  market regulations \cite{com2},
and transactions costs \cite{Davis2}. Analysis of empirical financial records indicates that the data can exhibit fat tails. This
feature has been observed in many different markets (see e.g. \cite{bor}
and the references therein). Such dynamics can be observed in emerging markets where the number of sellers and buyers is low. Also an interest rate often exhibits the feature of constant periods appearing - e.g. in US between $2002$ and $2017$ \cite{bdt}. 
In response to the empirical evidence of fat tails, $\alpha$-stable distribution as an alternative to the Gaussian
law was proposed \cite{AW13, AW12}.
The stable distribution has found many important applications, for
example in finance \cite{AW14}, physics \cite{AW16, AW15, AW17} and electrical engineering \cite{AW18}.
With increasing interest of fractional calculus and non-local
differential operators the family of fractional B-S equations
has emerged in the recent literature (see e.g. \cite{ja2} and references therein).

The subdiffusive B-S is the generalization of the classical B-S model to the cases, where the underlying assets
display characteristic periods in which they stay motionless. The standard B-S model does not take this phenomena into account because it assumes the asset is described by continuous Gaussian random walk. As a result of an option pricing for such underlying asset, the fair price provided by the B-S model is misestimated.
In order to describe this dynamics properly, the subdiffusive B-S model assumes that the underlying instrument is driven by $\alpha$-stable inverse subordinator \cite{sato}. Then the frequency of the constant periods appearing is dependent of subdiffusion parameter $\alpha\in(0,1)$. If $\alpha\rightarrow1$, the subdifussive B-S is reducing to the classical model.
Due to its practicality and simplicity, the standard B-S model
is one of the most widely used in option pricing. Although in contrast to the subdiffusive case it does not take into account the empirical property of the constant price periods in the underlying instrument dynamics.
In Figure $1$ we compare sample simulation of underlying asset in classical and subdiffusive market model. Even short stagnation of a market can not be simulated by standard B-S model.
As a generalization of the classical B-S model, its subdiffusive equivalent can be used in wide range of markets - including all cases where B-S can be applied.

In the case of standard B-S many numerical methods for American and barrier options have been proposed. They include such methods and techniques as: linear complementarity \cite{fast3,ja,fast2,fast4}, Bermudan approximation \cite{fast1,fast2}, penalty method \cite{for1, element1,for2}, repeated Richardson extrapolation \cite{nd2,nd3}, finite difference \cite{ja,nd1,compact}/element \cite{madi1,element1} and binomial/trinomial methods \cite{bin11,bin1}.
Also numerous analytical and semi analytical methods have been developed which are based mainly on asymptotic approximation formulas, method of lines, fast Fourier transform and integral equations. Despite their high numerical efficiency they can not be adapted to many applications e.g. the standard B-S with variable volatility (see e.g. \cite{nd2} and references therein).

Since the subdiffusive B-S model was proposed \cite{MM} many open problems still have remained unsolved.
One of them is the way of valuation American and Exotic options. In this paper we derive the Linear Complementarity Problem (LCP) system describing the fair price of an
American option in subdiffusive B-S model. We apply the weighted scheme of the FD method and the Longstaff-Schwartz method to solve the system numerically.
We compare both methods, moreover we show how to valuate numerially wide range of barrier options in subdiffusive B-S model using the FD approach. The paper is an extension of \cite{ja2} where the governing fractional differential equation for the European option in the subdiffusive B-S is derived and the weighted finite difference method is proposed. Moreover in \cite{ja2} the optimal value of weighting parameter $\theta$ for the wide class of fractional diffusion-type problems is found.

\section{Subdiffusive B-S model}
The evolution of the market is taking place up to time horizon $T$ and is contained in the probability space
$\left(\Omega, \mathcal{F} , \mathbb{P}\right)$.
Here, $\Omega$ is the sample space,  $\mathcal{F}$ is a filtration interpreted as the information about history of asset price and $\mathbb{P}$ is the “objective” probability measure.
The assumptions are the same as in  the classical B-S model \cite{ja} with the exception that we do not have to assume the market liquidity and that
the underlying instrument, instead of Geometric Brownian Motion (GBM), follows subdiffusive GBM \cite{MM}:
$$
\left\{ \begin{array}{ll}
Z_{\alpha}\left(t\right)=Z\left(S_{\alpha}(t)\right),\\
Z\left(0\right)=Z_{0},
\end{array}\right.$$
where $Z_{\alpha}\left(t\right)$ - the price of the underlying instrument, $Z(t)=Z\left(0\right)\exp \left(\mu t+\sigma B_{t}\right)$, $\mu$ - drift (constant), $\sigma$ - volatility (constant), $B_{t}$- Brownian motion, $S_{\alpha}(t)$ - the inverse $\alpha$-stable subordinator defined as $S_{\alpha}(t)=\inf(\tau>0: U_{\alpha}(\tau)>t)$ \cite{MM}, where $U_{\alpha}(t)$ is the $\alpha$-stable subordinator \cite{sato}, $0<\alpha<1$. Here $S_{\alpha}(t)$ is independent  of $B_{t}$ for each $t\in[0,T]$.
In Figure $1$ we compare the samples trajectories of GBM and subdifussive GBM for given $\alpha$-stable subordinator.
\begin{figure}[h]
\centering
\includegraphics[scale=0.6]{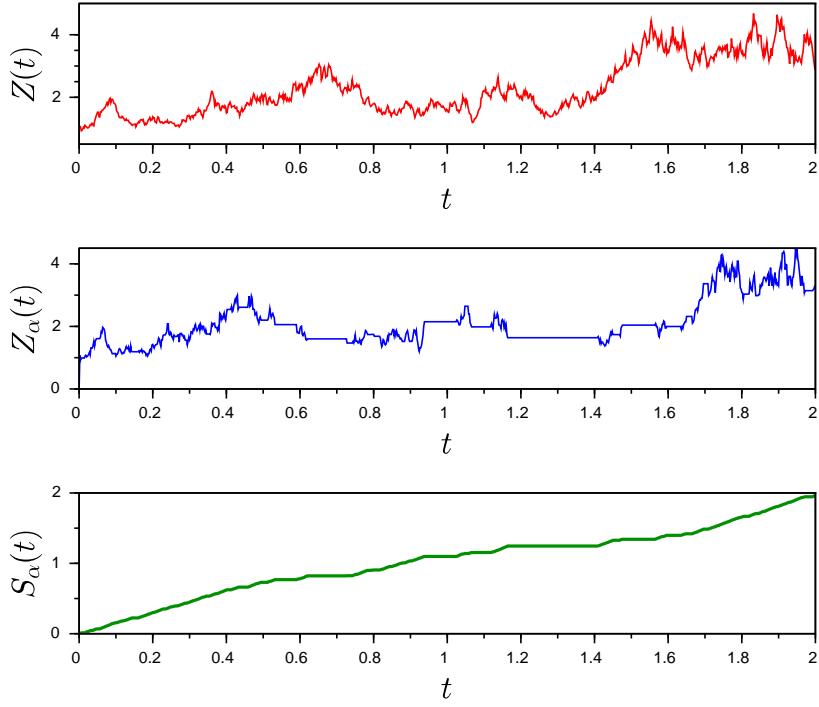}
\caption{The sample trajectory of GBM (upper panel) with its subdiffusive analogue (middle panel) and the corresponding inverse subordinator (lower panel). In the subdiffusive GBM the constant periods characteristic for
emerging markets can be observed. The parameters are $Z_{0}=\sigma=\mu=1$, $\alpha=0.9$.}
\end{figure}

 Let us introduce the probability measure \begin{equation}
                                            \mathbb{Q}\left(A\right)=\int_{A}\exp\left(-\gamma B\left(S_{\alpha}\left(T\right)\right)-\frac{\gamma^{2}}{2}S_{\alpha}\left(T\right)\right)dP,\label{1}
                                           \end{equation}
  where $\gamma=(\mu+\frac{\sigma^2}{2})/\sigma$, $A\in\mathcal{F}$. As it is shown in \cite{MM}, $Z_{\alpha}(t)$ is a $\mathbb{Q}$-martingale.
The subdiffusive B-S model is arbitrage-free and incomplete \cite{MM}. Despite $\mathbb{Q}$ defined in (\ref{1}) is not unique, but it is the ``best'' martingale
measure in the sense of criterion of minimal relative entropy. It means that the measure $\mathbb{Q}$ minimizes the distance to the measure $\mathbb{P}$ \cite{Gajda}. Between European put and call options the put-call parity holds \cite{MM}.

\section{Selected options}
In Tables \ref{table:1} and \ref{table:2} we recall the payoff functions for options considered in this article. Recall that the payoff function $f(Z_{t})$ is the gain of the option holder at the time $t$ for given underlying instrument $Z$.
\begin{table}[!ht]\footnotesize
\begin{center}
\begin{tabular}{|c c c|}
\hline

&European&American\\ \hline
Plain& $\max\left(Z_{T}-K,0\right)$& $\max\left(Z_{t}-K,0\right)$\\			
Knock up-and-in&$\max\left(Z_{T}-K,0\right)\mathds{1}{\{M_{T}>H^{+}\}}$& $\max\left(Z_{t}-K,0\right)\mathds{1}{\{M_{t}>H^{+}\}}$\\		
Knock up-and-out& $\max\left(Z_{T}-K,0\right)\mathds{1}{\{M_{T}<H^{+}\}}$& $\max\left(Z_{t}-K,0\right)\mathds{1}{\{M_{t}<H^{+}\}}$\\			
Knock down-and-in& $\max\left(Z_{T}-K,0\right)\mathds{1}{\{m_{T}<H^{-}\}}$& $\max\left(Z_{t}-K,0\right)\mathds{1}{\{m_{t}<H^{-}\}}$\\				
Knock down-and-out& $\max\left(Z_{T}-K,0\right)\mathds{1}{\{m_{T}>H^{-}\}}$&$\max\left(Z_{t}-K,0\right)\mathds{1}{\{m_{t}>H^{-}\}}$ \\	
Knock double-out& $\max\left(Z_{T}-K,0\right)\mathds{1}{\{H^{+}>M_{T},m_{T}>H^{-}\}}$& $\max\left(Z_{t}-K,0\right)\mathds{1}{\{H^{+}>M_{t},m_{t}>H^{-}\}}$\\	
Knock double-in& $\max\left(Z_{T}-K,0\right)-\max\left(Z_{T}-K,0\right)\mathds{1}{\{H^{+}>M_{T},m_{T}>H^{-}\}}$& $\max\left(Z_{t}-K,0\right)-\max\left(Z_{t}-K,0\right)\mathds{1}{\{H^{+}>M_{t},m_{t}>H^{-}\}}$\\	

\hline
\end{tabular}
\caption{\label{table:1}Payoff functions for selected call options.}
\end{center}
\end{table}

\begin{table}[!ht]\footnotesize
\begin{center}
\begin{tabular}{|c c c|}
\hline

&European&American\\ \hline
Plain&$\max\left(K-Z_{T},0\right)$& $\max\left(K-Z_{t},0\right)$\\			
Knock up-and-in&$\max\left(K-Z_{T},0\right)\mathds{1}{\{M_{T}>H^{+}\}}$& $\max\left(K-Z_{t},0\right)\mathds{1}{\{M_{t}>H^{+}\}}$\\		
Knock up-and-out&$\max\left(K-Z_{T},0\right)\mathds{1}{\{M_{T}<H^{+}\}}$& $\max\left(K-Z_{t},0\right)\mathds{1}{\{M_{t}<H^{+}\}}$\\			
Knock down-and-in&$\max\left(K-Z_{T},0\right)\mathds{1}{\{m_{T}<H^{-}\}}$& $\max\left(K-Z_{t},0\right)\mathds{1}{\{m_{t}<H^{-}\}}$\\				
Knock down-and-out&$\max\left(K-Z_{T},0\right)\mathds{1}{\{m_{T}>H^{-}\}}$& $\max\left(K-Z_{t},0\right)\mathds{1}{\{m_{t}>H^{-}\}}$\\	
Knock double-out&$\max\left(K-Z_{T},0\right)\mathds{1}{\{H^{+}>M_{T},m_{T}>H^{-}\}}$& $\max\left(K-Z_{t},0\right)\mathds{1}{\{H^{+}>M_{t},m_{t}>H^{-}\}}$\\	
Knock double-in&$\max\left(K-Z_{T},0\right)-\max\left(K-Z_{T},0\right)\mathds{1}{\{H^{+}>M_{T},m_{T}>H^{-}\}}$& $\max\left(K-Z_{t},0\right)-\max\left(K-Z_{t},0\right)\mathds{1}{\{H^{+}>M_{t},m_{t}>H^{-}\}}$\\	

\hline
\end{tabular}
\caption{\label{table:2}Payoff functions for selected put options.}
\end{center}
\end{table}
Here and in the rest of the paper $K$ - strike, $Z_{t}=Z_{\alpha}(t)$ - value of underlying instrument at time $t$, $t\in[0,T]$, $\displaystyle M_{t}=\max_{\tau\in[0,t]}\left(Z_{\tau}\right)$,
$\displaystyle m_{t}=\min_{\tau\in[0,t]}\left(Z_{\tau}\right)$, $H^{+}$, $H^{-}$ - barriers.

\section{Valuation of American option as free boundary problem}
In whole paper we assume that the dividend rate $\delta=0$. The next proposition explains why in context of American options we will proceed only with the put options.
\begin{Proposition}
If the dividend rate $\delta=0$ and $r\geq0$, then the value of American call option is equal to its
 European analogue.
 Similarly it can be shown that if $r=0$, then it is not worth to realize American put before $T$, so in this case value of American put is equal to his European equivalent.
\end{Proposition}
\begin{Proof}\cite{ja}
 Let us assume $\delta=0$, $r\geq0$ and that the American call option is intended to be realized at time $t<T$. It will be done only if it would be profitable, it means if $Z_{t}>K$. Then, the expected profit (i.e. the future expected profit computed at time $t=0$) is $(Z_{t}-K)e^{-rt}$.
Option holder can also proceed following strategy: at time $t$ short sell an underlying instrument for $Z_{t}$, next buy it (using the American or European option) for $\min\left(Z_{T},K\right)$ at time $T$. Then his expected profit is $Z_{t}e^{-rt}-\min\left(Z_{T},K\right)e^{-rT}$. Let us observe that the following sequence of inequalities holds:\[Z_{t}e^{-rt}-\min\left(Z_{T},K\right)e^{-rT}\geq \left(Z_{t}-\min\left(Z_{T},K\right)\right)e^{-rt}\geq (Z_{t}-K)e^{-rt}.\] Hence, earlier realization of the American option is
at most as profitable as use of the option at $t = T$. If the gain of American option holder in the most profitable strategy is the same as the profit of European option holder, we conclude that both instruments have the same value. Indeed, the fair price of an option is equal to the expected gain of the most profitable strategy provided by the option. Analogously it can be shown the second statement.
\end{Proof}

We proceed with the following main result of this section

\begin{Theorem}
 The fair price of an American put option in the subdiffusive B-S model is equal to $v(z, t)$, where $v(z, t)$ satisfies:
\begin{equation}
\begin{cases}
 x=\ln z,\\
u\left(x,t \right)=v\left(e^{x},T-t \right).
\end{cases}\label{1b}
\end{equation}
and $u(x, t)$ is the solution of the system
\begin{equation}
\begin{cases}
   u\left(x,0\right)=\max\left(K-\exp\left(x\right),0\right),\\
 u\left(x,t\right)\geq \max\left(K-\exp\left(x\right),0\right)\text{, }t\in[0,T]\\
  {}_{0}^{c}D^{\alpha}u\left(x,t\right)-\displaystyle\frac{1}{2}\sigma^2\displaystyle\frac{\partial^2 u\left(x,t\right)}{\partial x^2}-\left(r-\displaystyle\frac{1}{2}\sigma^2\right)\displaystyle\frac{\partial u\left(x,t\right)}{\partial x}+ru\left(x,t\right)\geq 0\text{, }t\in(0,T)\\
    \left(u\left(x,t\right)-\max\left(K-\exp\left(x\right),0\right)\right)\cdot \left( {}_{0}^{c}D^{\alpha}u\left(x,t\right)-\displaystyle\frac{1}{2}\sigma^2\displaystyle\frac{\partial^2 u\left(x,t\right)}{\partial x^2}-\left(r-\displaystyle\frac{1}{2}\sigma^2\right)\displaystyle\frac{\partial u\left(x,t\right)}{\partial x}+ru\left(x,t\right)\right)=0\text{, }t\in(0,T)\\
\lim\limits_{x\rightarrow\infty}u\left(x,t\right)=0,\\
 \lim\limits_{x\rightarrow -\infty}u\left(x,t\right)=K,\\
 \end{cases}\label{2}
\end{equation}
where $x\in\left(-\infty,\infty\right)$ and ${}_{0}^{c}D_{t}^{\alpha}$ is Caputo fractional derivative defined as \cite{com3}:
\[{}_{0}^{c}D_{t}^{\alpha}g\left(t\right)=\frac{1}{\Gamma\left(1-\alpha\right)}\int_{0}^{t}\frac{d g\left(s\right)}{d s}\left(t-s\right)^{-\alpha}ds.\]

\end{Theorem}

\begin{Proof}
We consider the subdiffusive B-S Equation \cite{ja2},
\begin{equation}
\displaystyle {}_{0}^{c}D_{t}^{\alpha}v\left(z,t\right)=
 -\frac{1}{2}\sigma^2z^{2}\frac{\partial^2 v\left(z,t\right)}{\partial z^2}-rz\frac{\partial v\left(z,t\right)}{\partial z}+rv\left(z,t\right)\text{,  } (z,t)\in\left(0,\infty\right)\times(0,T),\label{3}
 \end{equation}
 with the terminal condition determining put option

 \begin{equation}v\left(z,T\right)=\max\left(K-z,0\right)\text{,  } z\in\left(0,\infty\right).\label{4} \end{equation}
At the time $t\in[0,T]$ we can gain at least $\max\left(K-z,0\right)$ (by exercising the option) and maybe even more. It leads us to the inequality:\newline
\begin{equation}
 v\left(z,t\right)\geq \max\left(K-z,0\right).\label{5}
\end{equation}
After the optimal exercise moment $v\left(z,t\right)$ can not describe the value process of the optimal strategy (because we "missed" the best moment to use the option). Hence, we can gain at most the value provided by keeping the option - i.e. by the solution of (\ref{3}). In other words, the differential equation (\ref{3}) can underestimate the value of the option. So for $t\in(0,T)$ true is the following inequality:

\begin{equation}
{}_{0}^{c}D_{t}^{\alpha}v\left(z,t\right)+\displaystyle\frac{1}{2}\sigma^2 z^{2}\displaystyle\frac{\partial^2 v\left(z,t\right)}{\partial z^2}\leq rv\left(z,t \right) - rz\displaystyle\frac{\partial v\left(z,t\right)}{\partial z}.\label{6}
\end{equation}
Indeed, the differential inequality (\ref{6}) cannot formally be interpreted that over any infinitesimal time interval the sum of the losses and profits from holding the option is lower or equal to the return at the riskless rate. Note that the time derivative from (\ref{6}) can be interpreted as loss in value due to having less time for exercising the option and the derivative of second order over the price of the underlying instrument can be interpreted as the profit in holding the option.
\\At each moment $t\in(0,T)$ we decide if it is worth to use the option, or keep it.
Mathematically we can describe it as:
 $$v\left(z,t\right)=\max\left(K-z,0\right),$$ if we realize option or $$ \displaystyle {}_{0}^{c}D_{t}^{\alpha}v\left(z,t\right)
 +\frac{1}{2}\sigma^2z^{2}\frac{\partial^2 v\left(z,t\right)}{\partial z^2}+rz\frac{\partial v\left(z,t\right)}{\partial z}-rv\left(z,t\right)=0,
 $$ if we keep it on. This can be written as follows:
 \begin{equation}
 \left(v\left(z,t\right)-\max\left(K-z,0\right)\right)\cdot \left( {}_{0}^{c}D_{t}^{\alpha}v\left(z,t\right)
 +\frac{1}{2}\sigma^2z^{2}\frac{\partial^2 v\left(z,t\right)}{\partial z^2}+rz\frac{\partial v\left(z,t\right)}{\partial z}-rv\left(z,t\right)\right)=0.\label{7}
  \end{equation}
For the sufficiently high price of the underlying instrument we will not use the option. So:
\begin{equation}
\lim\limits_{z\rightarrow\infty}v\left(z,t\right)=0.\label{8}
\end{equation}
In analogy if the price of the underlying instrument will be low we will use the option, selling the underlying instrument
for $K$:
\begin{equation}
\lim\limits_{z\rightarrow 0}v\left(z,t\right)=K.\label{9}
\end{equation}

After the change of variables (\ref{1b}), (\ref{4})-(\ref{9}) have the form of (\ref{2}).
\end{Proof}


\section{Numerical scheme for American put option}
In this section we derive the numerical scheme for American put option. To do so, we will approximate limits by finite numbers and derivatives by finite differences.  We will proceed for a $\theta$-convex combination
of explicit ($\theta=1$) and implicit ($\theta=0$) discrete scheme, similarly as it was done for European options in \cite{ja2}. In the same work the corresponding stability/convergence analysis can be found. We introduce parameter $\theta\in[0,1]$ by optimization purposes - similarly as for the case $\alpha=1$, $\theta=\frac{1}{2}$ has the best properties in terms of the error and unconditional stability/convergence \cite{ja}.
Instead of the continuous space $\left(-\infty,\infty\right)\times [0,T]$, we take its discrete and finite equivalent
$\{x_{0}=x_{min},x_{1}\ldots, x_{n}=x_{max}\}\times\{t_{0}=0,\ldots,t_{N}=T\}$, where $x_{min}$, $x_{max}$ - lower and upper boundary of the grid.
We consider a uniform grid, so $t_{j}=j\Delta t$  and $x_{i}=x_{min}+i\Delta x$, where $i=0,\ldots,n$, $j=0,\ldots,N$, $\Delta t=T/N$, $\displaystyle  \Delta x=(x_{max}-x_{min})/n$.
After obtaining the discrete analogue of (\ref{2}) we will solve it recursively. As a result we will find its numerical solution $\hat{u}_{i}^{j}=\hat{u}(x_{i},t_{j})$, $i=0,\ldots,n$, $j=0,\ldots,N$.

We begin discretizing initial condition for put option (first line of (\ref{2})), we get
\begin{equation}
 \hat{u}^{0}_{i}=\max\left(K-\exp\left(x_{min}+i\Delta x\right),0\right),\label{10}
\end{equation}
for $i=0,\ldots,n$.\\ Similarly the discrete version of boundary conditions (the last $2$ lines of (\ref{2})) has the form
\begin{equation}
\begin{cases}
\hat{u}^{l}_{n}=0,\\
\hat{u}^{l}_{0}=K,\\
\end{cases}\label{11}
\end{equation}
for $l=0,\ldots,N$.\\ Discretizing third, second and fourth lines of (\ref{2}) we get
\begin{equation}
\begin{cases}
C\hat{u}^{1}=\hat{u}^{0}+\left(1-\theta\right)G^{1} +\theta G^{0}+\theta B\hat{u}^{0},\\
\hat{u}^{1}=\max(\hat{u}^{1},\hat{u}^{0}),\\
C\hat{u}^{k+1}=\displaystyle\sum_{j=0}^{k-1}\left(b_{j}-b_{j+1}\right)\hat{u}^{k-j}+b_{k}\hat{u}^{0}+\left(1-\theta\right)G^{k+1} +\theta G^{k}+\theta B\hat{u}^{k},\\
\hat{u}^{k+1}=\max\left(\hat{u}^{k+1},\hat{u}^{0}\right),
\end{cases}\label{12}
\end{equation}
for $k\geq1$.\\

Here $b_{j}=\left(j+1\right)^{1-\alpha}-j^{1-\alpha},$
$C=\left(\theta I+\left(1-\theta\right) A\right),$
$A=\left(a_{ij}\right)_{\left(n-1\right)\times\left(n-1\right)}$, such  that:\newline
$a_{ij}=$
 $
\begin{cases}
1+2\displaystyle\frac{ad}{\Delta x^2}+cd,& \text{for } j=i,i=1,2...,n-1,\\
-\left(\displaystyle\frac{ad}{\Delta x^2}-\displaystyle\frac{bd}{2\Delta x}\right),& \text{for } j=i-1,i=2...,n-1,\\
-\left(\displaystyle\frac{ad}{\Delta x^2}+\displaystyle\frac{bd}{2\Delta x}\right),& \text{for } j=i+1,i=2...,n-2,\\
0,& \text{in other cases,}
\end{cases}
$\newline
\newline$B=\left(b_{ij}\right)_{\left(n-1\right)\times\left(n-1\right)}$, such  that:\newline
$b_{ij}=$
 $
\begin{cases}
-\left(2\displaystyle\frac{ad}{\Delta x^2}+cd\right),& \text{for } j=i,i=1...,n-1,\\
\displaystyle\frac{ad}{\Delta x^2}+\displaystyle\frac{bd}{2\Delta x},& \text{for } j=i+1,i=1,2...,n-1,\\
\displaystyle\frac{ad}{\Delta x^2}-\displaystyle\frac{bd}{2\Delta x},& \text{for } j=i-1,i=1,2...,n-2,\\
0,& \text{in other cases,}
\end{cases}
$
$$G^{k}=\left(\left(\displaystyle\frac{ad}{\Delta x^2}-\displaystyle\frac{bd}{2\Delta x}\right)\hat{u}^{k}_{0},0,...,0,\left(\displaystyle\frac{ad}{\Delta x^2}+\displaystyle\frac{bd}{2\Delta x}\right)\hat{u}^{k}_{n-1}\right)^{T},$$
$$u^{k}=\left(u_{1}^{k},u_{2}^{k},...,u_{n-1}^{k}\right)^{T},$$
$a=\displaystyle\frac{1}{2}\sigma^2$, $b=\left(r-\displaystyle\frac{1}{2}\sigma^2\right)$, $c=r$,
$d=\Gamma\left(2-\alpha\right)\Delta t^{\alpha},$
$\Delta t=T/N,$ $k=1,2,...N.$\\ Note that the analogical scheme for the European option \cite{ja2} is
\begin{equation}
\begin{cases}
C\hat{u}^{1}=\hat{u}^{0}+\left(1-\theta\right)G^{1} +\theta G^{0}+\theta B\hat{u}^{0},\\
C\hat{u}^{k+1}=\displaystyle\sum_{j=0}^{k-1}\left(b_{j}-b_{j+1}\right)\hat{u}^{k-j}+b_{k}\hat{u}^{0}+\left(1-\theta\right)G^{k+1} +\theta G^{k}+\theta B\hat{u}^{k},
\end{cases}\label{13}
\end{equation}
with corresponding boundary conditions
\begin{equation}
\begin{cases}
\hat{u}^{l}_{n}=\exp(x_{max})-K\exp(-r\left(T-t_{l}\right)),\\
\hat{u}^{l}_{0}=0,\\
\end{cases}\label{14}
\end{equation}
 and initial condition for a call option
 \begin{equation}
  \hat{u}^{0}_{i}=\max\left(\exp\left(x_{min}+i\Delta x\right)-K,0\right),\label{15}
 \end{equation}
where $l=0,\ldots,N$, $i=0,1\ldots,n$.
\begin{figure}[ht]
\centering
\includegraphics[scale=0.59]{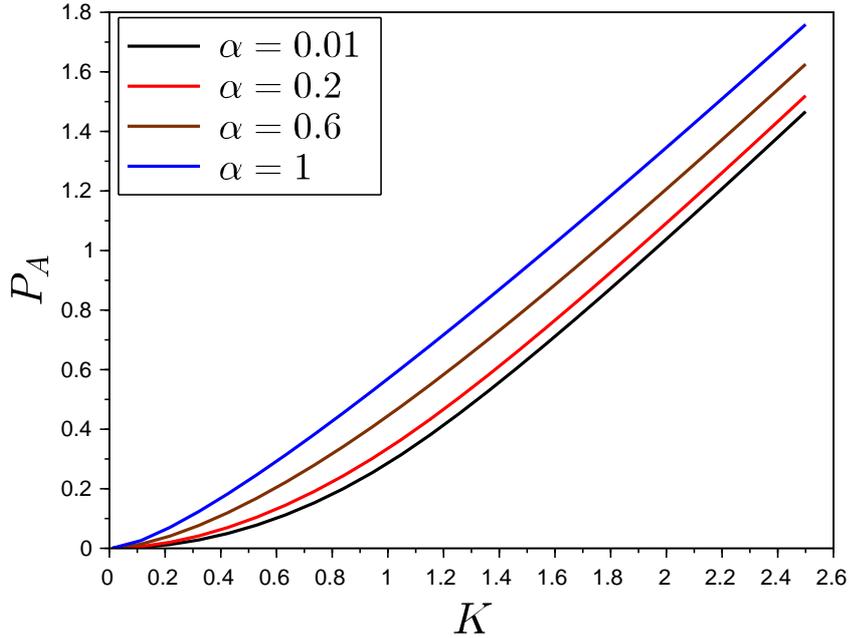}
\caption{The price of American put ($P_{A}$) in dependence of $K$. The parameters are $n=1000$, $x_{min}=-20$, $x_{max}=10$
$\sigma=1$, $r=0,04$, $N=140$, $T=4$, $Z_{0}=1$, $\theta=\check{\theta_{\alpha}}$.}
\end{figure}

\section{Numerical schemes for barrier options}

The systems (\ref{12}) and (\ref{13}) can be used to price different types of options, only if the initial-boundary conditions will be properly modified.
Let us treat $x_{min}$ and $x_{max}$ not as approximations of infinite values, but as logarithm of lower and supreme barriers $H^{-}$ and $H^{+}$ defined in double barrier option. We take a logarithm because of change of variables $x=\ln{z}$ made in (\ref{1b}) and in \cite{ja2}.
The initial
 \begin{equation}
  \hat{u}^{0}_{i}=\max\left(\exp\left(\ln{H^{-}}+i\Delta x\right)-K,0\right),\label{16}
 \end{equation}
 and boundary conditions
\begin{equation}
\begin{cases}
\hat{u}^{l}_{n}=0,\\
\hat{u}^{l}_{0}=0,\\
\end{cases}\label{17}
\end{equation}
where
$l=0,1\ldots,N$, $\Delta x=(\ln{H^{+}}-\ln{H^{-}})/n$, $i=0,1\ldots,n,$\\
together with (\ref{13}) is the scheme for the European double knock-out call option. The same boundary-initial conditions with (\ref{12}) create the scheme for the American double knock-out call option.
Analogously, prices of one-side barrier knock-out options can be obtained. Hence we have initial and boundary conditions for knock-down-and-out call option
\begin{equation}
  \hat{u}^{0}_{i}=\max\left(\exp\left({\ln{H^{-}}}+i\Delta x\right)-K,0\right),\label{18}
\end{equation}

\begin{equation}
\begin{cases}
\hat{u}^{l}_{n}=\exp(x_{max})-K\exp(-r\left(T-t_{l}\right)),\\
\hat{u}^{l}_{0}=0,
\end{cases}\label{19}
\end{equation}
$l=0,\ldots,N$, $\Delta x=(x_{max}-\ln{H^{-}})/n$, $i=0,1\ldots,n$,

and for knock-up-and-out call option

\begin{equation}
  \hat{u}^{0}_{i}=\max\left(\exp\left(x_{min}+i\Delta x\right)-K,0\right),\label{20}
\end{equation}

\begin{equation}
\begin{cases}
\hat{u}^{l}_{0}=0,\\
\hat{u}^{l}_{n}=0,
\end{cases}\label{21}
\end{equation}
$l=0,\ldots,N$, $\Delta x=(\ln{H^{+}}-x_{min})/n$, $i=0,1\ldots,n$.

\begin{figure}[ht]
\centering
\includegraphics[scale=0.55]{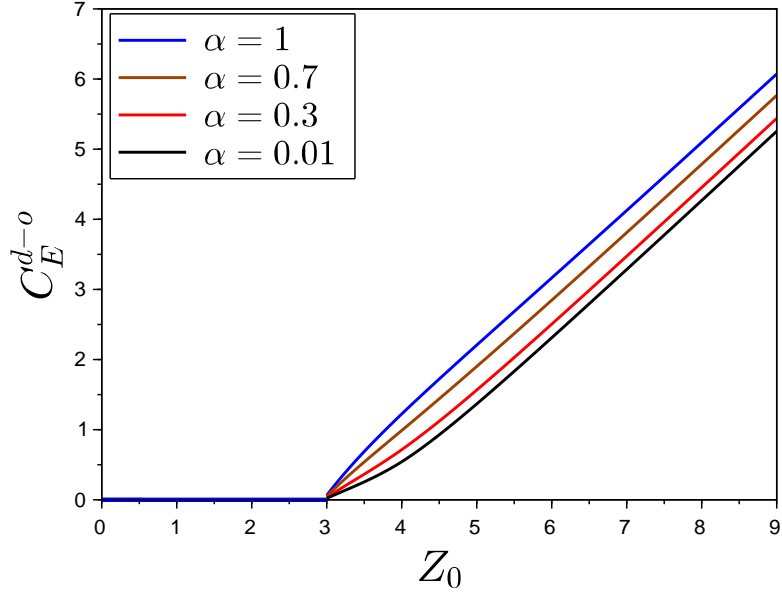}
\caption{The price of European down-and-out call option ($C_{E}^{d-o}$) in dependence of $Z_{0}$. The parameters are $n=300$,
$\sigma=0,3$,
$r=0,08$,
$N=300$,
$x_{max}=100$,
$H^{-}=3$,
$T=4$,
$K=4$,
$\theta=\check{\theta_{\alpha}}$.}
\end{figure}
If we want to price the knock-in options, it is helpful to use the fact that for fixed parameters there holds the so called in-out parity
\begin{equation*}
 Knock_{in}=Van-Knock_{out},
\end{equation*}
where $Van$ - the price of Vanilla (plain) option, $Knock_{in}, Knock_{out}$ - option prices of knock-in and knock-out of the same type and style.\\

Please note, that the value of a double knock-out option for $Z_{0}$ outside of the interval $\left(\ln{H^{-}},\ln{H^{+}}\right)$ (but being a positive number) is equal $0$.  Analogous remark applies for one-sided barrier options.

To summarize we present the way to price the considered options in Tables \ref{table:3}, \ref{table:4} and \ref{table:5}.

\begin{figure}[ht]
\centering
\makebox[\textwidth][c]{\includegraphics[width=1.24\textwidth]{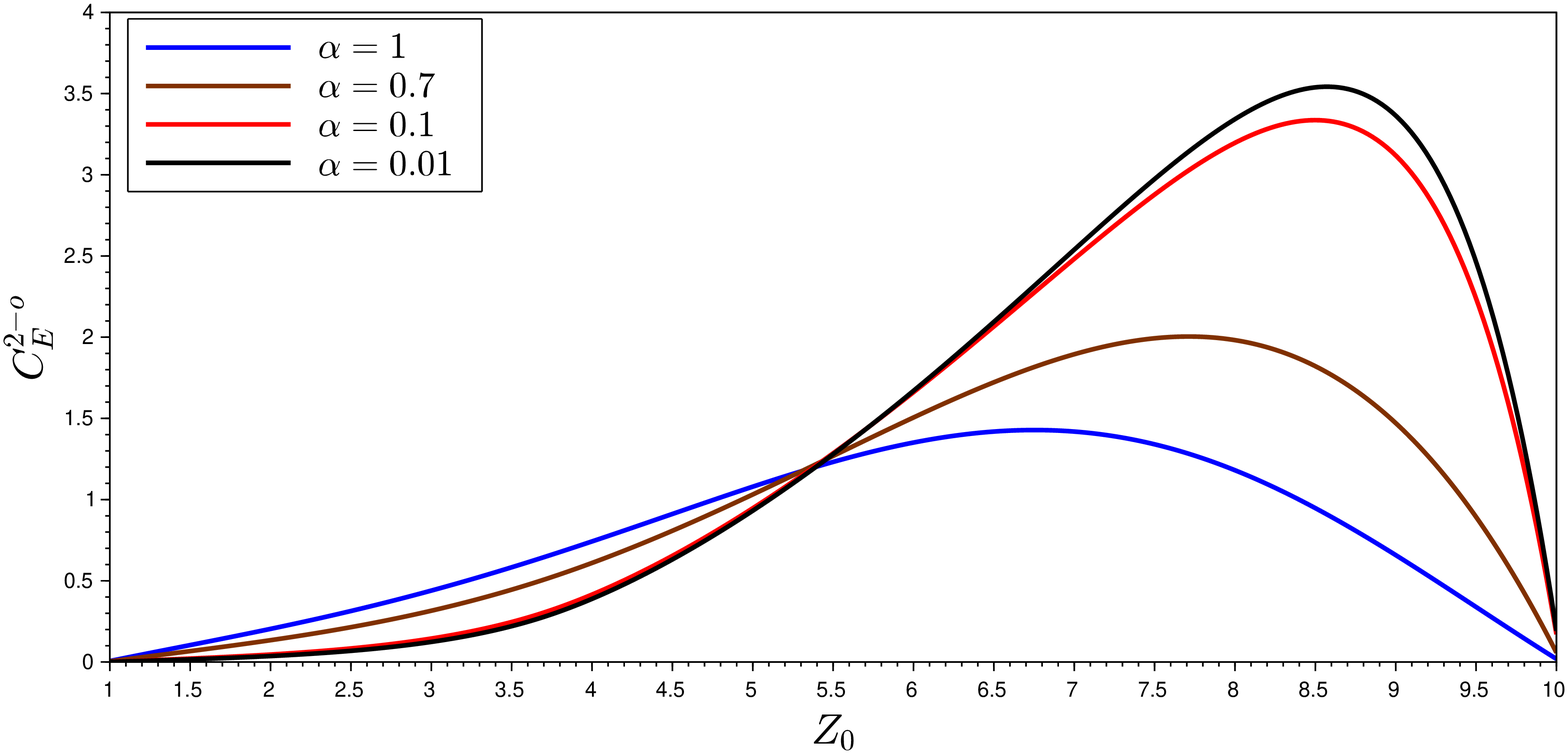}}
\caption{The price of European double knock-out call option ($C_{E}^{2-o}$) in dependence of $Z_{0}$. The parameters are $n=300$, $\sigma=0,3$,
$r=0,08$,
$N=300$,
$H^{+}=10$,
$H^{-}=1$,
$T=4$,
$K=2$,
$\theta=0$.
In this case, there is no clear relation between the prices of option for different $\alpha$ (like e.g. in vanilla equivalent for $T>1$ where for higher $\alpha$ the price is higher for all $Z_{0}$).
We can conclude that in this figure there is a critical point where all plots intersect. It is unknown under which conditions (if there is any) such point exists and what is its value. }
\end{figure}\newpage

\begin{table}[!ht]\footnotesize
\begin{center}
\begin{tabular}{|c c c c c|}
\hline

Style of option&Numerical scheme&Boundary conditions&Initial condition&Apply in-out parity?\\ \hline
Plain& (\ref{13})& (\ref{14})&(\ref{15})&No\\			
Knock up-and-in& (\ref{13})& (\ref{21})&(\ref{20})&Yes\\
Knock up-and-out& (\ref{13})& (\ref{21})&(\ref{20})&No\\		
Knock down-and-in& (\ref{13})& (\ref{19})&(\ref{18})&Yes\\		
Knock down-and-out& (\ref{13})& (\ref{19})&(\ref{18})&No\\	
Knock double-out& (\ref{13})& (\ref{17})&(\ref{16})&No\\
Knock double-in& (\ref{13})& (\ref{17})&(\ref{16})&Yes\\
\hline
\end{tabular}
\caption{\label{table:3}European call options.}
\end{center}
\end{table}

To price the European put options we can firstly compute their call equivalents and then apply the Put-Call parity. We can also use other initial conditions than in Table \ref{table:3},

\begin{table}[!ht]\footnotesize
\begin{center}
\begin{tabular}{|c c c c c|}
\hline

Style of option&Numerical scheme&Boundary conditions&Initial condition&Apply in-out parity?\\ \hline
Plain& (\ref{13})& (\ref{11})&(\ref{10})&No\\			
Knock up-and-in& (\ref{13})& (\ref{21put})&(\ref{25})&Yes\\
Knock up-and-out& (\ref{13})& (\ref{21put})&(\ref{25})&No\\		
Knock down-and-in& (\ref{13})& (\ref{19put})&(\ref{24})&Yes\\		
Knock down-and-out& (\ref{13})& (\ref{19put})&(\ref{24})&No\\	
Knock double-out& (\ref{13})& (\ref{17})&(\ref{23})&No\\
Knock double-in& (\ref{13})& (\ref{17})&(\ref{23})&Yes\\
\hline
\end{tabular}
\caption{\label{table:4}European put options.}
\end{center}
\end{table}
where
\begin{equation}
 \hat{u}^{0}_{i}=\max\left(K-\exp\left(\ln{H^{-}}+i\Delta x\right),0\right),\label{23}
\end{equation}
for $\Delta x=\left(\ln{H^{+}}-\ln{H^{-}}\right)/n$ (double options),

\begin{equation}
 \hat{u}^{0}_{i}=\max\left(K-\exp\left(\ln{H^{-}}+i\Delta x\right),0\right),\label{24}
\end{equation}
\begin{equation}
\begin{cases}
\hat{u}^{l}_{n}=0,\\
\hat{u}^{l}_{0}=0,
\end{cases}\label{19put}
\end{equation}
for $l=0,\ldots,N$, $\Delta x=(x_{max}-\ln{H^{-}})/n$, $i=0,1\ldots,n$, (knock-down options),

\begin{equation}
 \hat{u}^{0}_{i}=\max\left(K-\exp\left(x_{min}+i\Delta x\right),0\right),\label{25}
\end{equation}

\begin{equation}
\begin{cases}
\hat{u}^{l}_{0}=K,\\
\hat{u}^{l}_{n}=0,
\end{cases}\label{21put}
\end{equation}
for $l=0,\ldots,N$, $\Delta x=(\ln{H^{+}}-x_{min})/n$, $i=0,1\ldots,n$ (knock-up options).

 If there is no dividend (the case we consider in this paper), the American call is equal its European equivalent so Table \ref{table:3}
 holds also for American call options. 
 For American put options we have,

\begin{table}[!ht]\footnotesize
\begin{center}
\begin{tabular}{|c c c c c|}
\hline

Style of option&Numerical scheme&Boundary conditions&Initial condition&Apply in-out parity?\\ \hline
Plain& (\ref{12})& (\ref{11})&(\ref{10})&No\\			
Knock up-and-in& (\ref{12})& (\ref{21put})&(\ref{25})&Yes\\
Knock up-and-out& (\ref{12})& (\ref{21put})&(\ref{25})&No\\		
Knock down-and-in& (\ref{12})& (\ref{19put})&(\ref{24})&Yes\\		
Knock down-and-out& (\ref{12})& (\ref{19put})&(\ref{24})&No\\	
Knock double-out& (\ref{12})& (\ref{17})&(\ref{23})&No\\
Knock double-in& (\ref{12})& (\ref{17})&(\ref{23})&Yes\\
\hline
\end{tabular}
\caption{\label{table:5}American put options.}
\end{center}
\end{table}
\newpage
Note, that for each type of barrier option, the definition of $\Delta x$ is different.

In Figures $2$, $3$,$4$ we compare the fair prices given for different values of $\alpha$ for American put, European down-and-out call and European double knock-out call option respectively.

We recall the observation from \cite{ja2} that the optimal choice of $\theta$ for given $\alpha$
is such that $\displaystyle\check{\theta_{\alpha}}=\frac{2-2^{1-\alpha}}{3-2^{1-\alpha}}$.
Then the lowest boundary for an error is achieved without losing the unconditional stability/convergence. For $\displaystyle\theta>\check{\theta_{\alpha}}$
the stability/convergence is not provided. In Figure $5$ the relation between the fair price of American put $P_{A}$ and $\theta$ is presented. The real price of the option
is close to $0,25$. The jump presented in the figure is the result of the increasing error. It is the consequence of lack of the stability \cite{ja2}.

\begin{figure}[ht]
\centering
\includegraphics[scale=0.47]{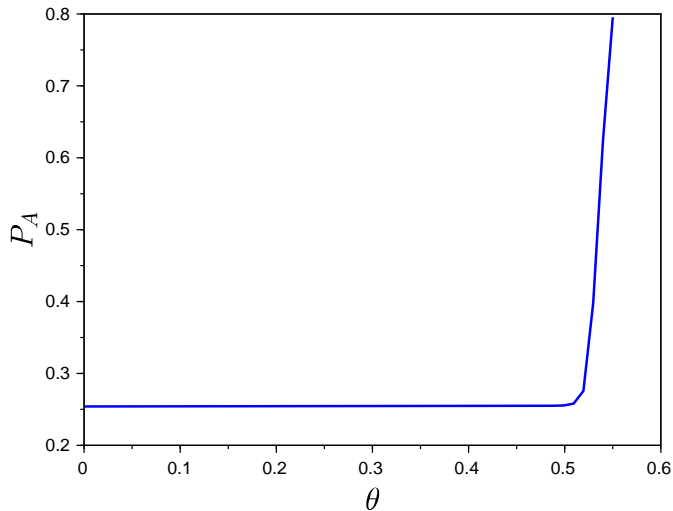}
\caption{The price of American put ($P_{A}$) in dependence of $\theta$. The explosion of the numerical error is the consequence of the lack of the unconditional stability outside of the interval $[0,\check{\theta_{\alpha}}]$ \cite{ja2}. The parameters are $n=1000$,
$\sigma=0,5$, $r=0,04$, $N=100$, $T=4$, $Z_{0}=1$, $K=1$, $\alpha=0,7$, $x_{min}=-20$, $x_{max}=10$.}
\end{figure}

\subsection{Longstaff-Schwartz method}
The Longstaff-Schwartz (LS) method is one of the most popular approaches for valuing American/Bermudan options and their Asian equivalents \cite{corn, LS}. Moreover it has an important applications in solving dynamic investment portfolio problems and in American/Bermuda style swaptions valuation (see e.g. \cite{LSswap} and references therein).
All these applications have an important meaning in finance. Only for notional amount of interest rate
swaps outstanding at the end of $1999$ the losses caused by wrong exercise strategies were estimated on billions of dollars \cite{LSswap}.

The main idea of this method is the use of least squares to
estimate the conditional expected payoff to the option holder from continuation. This
strategy allows to find the value of an option with the optimal exercise time (i.e. the moment where exercising option is the most profitable,
if there are more than one such moments we choose the lowest of them). The method was introduced for the classical B-S model but it can be extended for many other cases \cite{LS}.
In this paper we focus on American option, but note that the same method can be used to price Bermudan and American-style Asian options.
The LS algorithm can be found in Appendix.

Note that the inverse-$\alpha$ stable process is not Markovian so the expected value in $A.1$ could be taken not only under current but also previous states \cite{LS}.
Such proceeding could increase precision but will cost significant gain of running time. We decided to simplify the algorithm considering the expected value in $A.1$ only by the current state.
Interesting could be problem of optimal choice of the set of states - e.g. using statistical background.

The LS method has its limitations. As it is shown in \cite{niestabLS}, for continuous underlying process and small values of $T$ the method is unstable. The reason is the ill-condition of the underlying regression problem \cite{niestabLS}.
The analytical formulas indicating regime where the method is stable and where is not are still unknown. This fact limitates the range of possible applications of the algorithm.
Note that the error in LS method could be of different origin. The first is produced by discretization the continuous stochastic processes into $m$ nodes and assumption that the American option can be exercise only at these points.
The second is related to Monte Carlo  method i.e. that we estimate the expected value by the mean of size $M$. The next possible origin of the error is coming from approximation of
the conditional expected value $A.1$ by the average of $l$ basis functions. The last possible type of the error is produced for non-Markovian underlying processes. Since these processes
have memory, the expected value $A.1$ should be conditioned not only under the current state but also under the whole history of the underlying asset.
Since in the algorithm the stochastic process is considered only at discrete nodes, even conditioning by current and all previous states produce an error.
In contrast to LS, the FD method produces an error coming only from discretization of the variables (and approximation of infinities by $x_{min}$ and $x_{max}$), therefore its stability/convergence is easier to analyse.

\subsection{Numerical examples}

\begin{Example}
We compare both methods presented in this paper in pricing American put option. Simulations are made for $r=0,04$, $Z_{0}=5$, $K=2$, $N=150$,
$x_{max}=10$, $x_{min}=-20$, $n=200$, $\sigma=1$, $m=100$, $M=3000$ and different values of $T$ and $\alpha$.
\end{Example}

\begin{figure}[ht]
\centering
\includegraphics[scale=0.32]{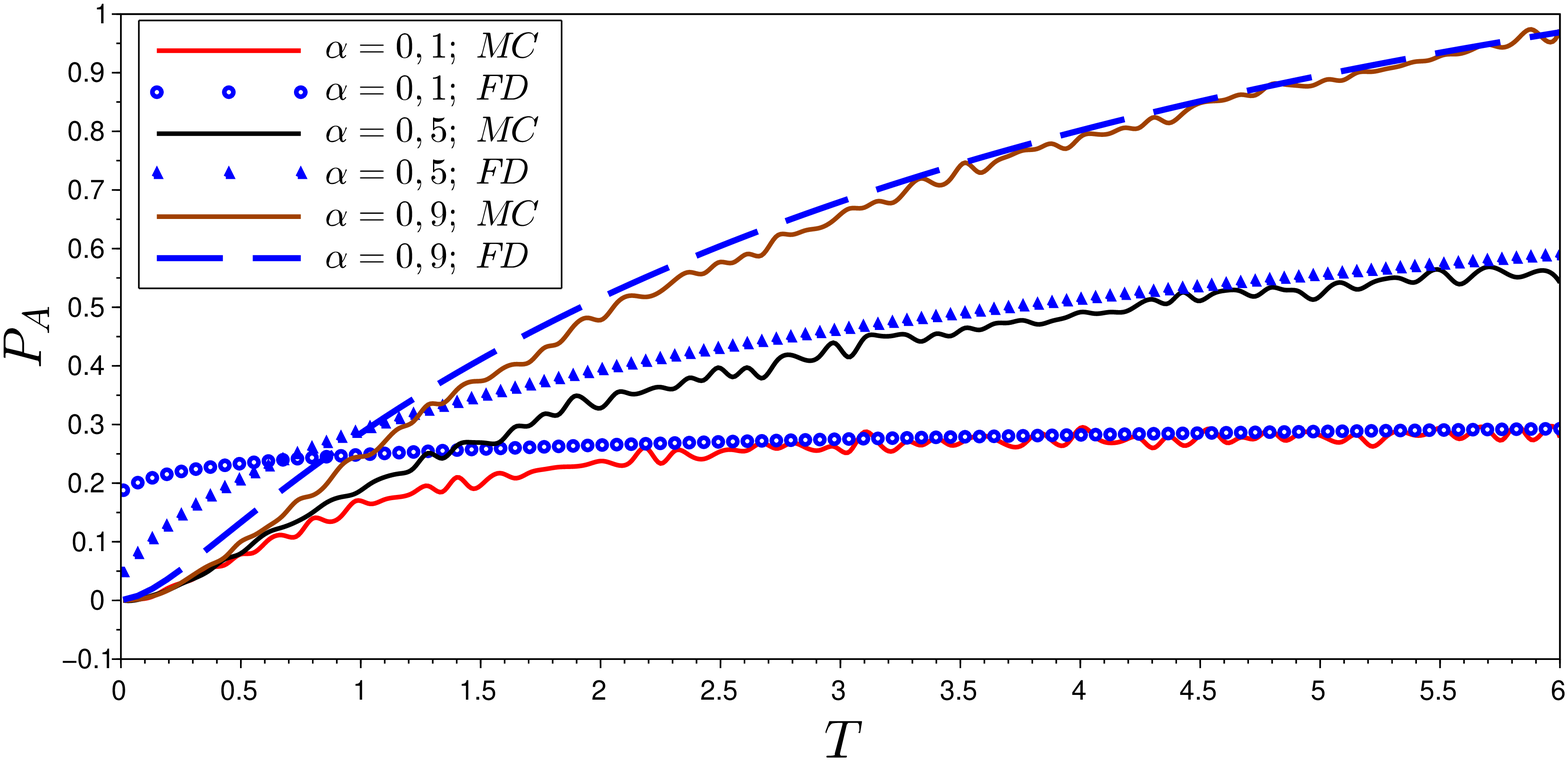}

\centering

\includegraphics[scale=0.3]{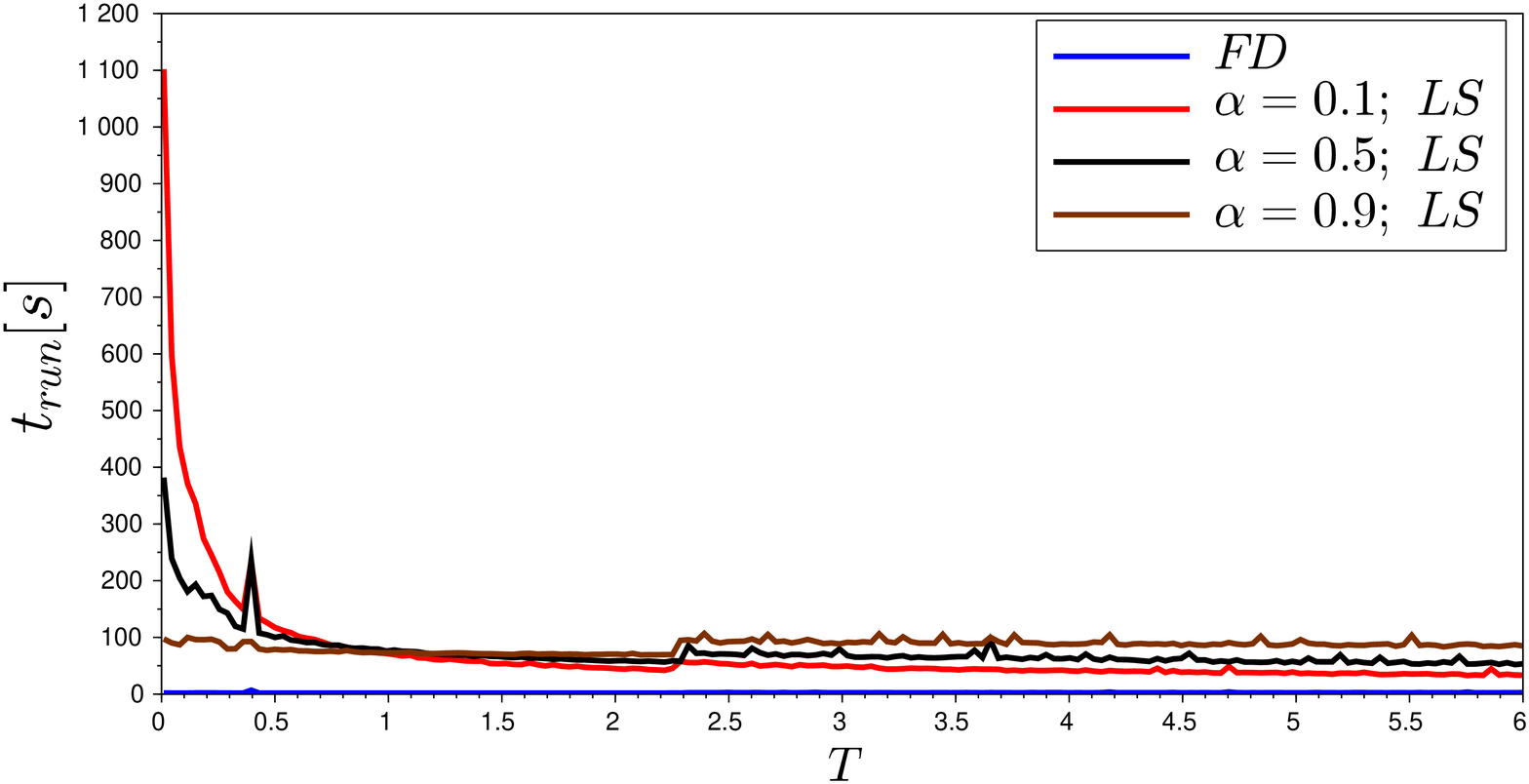}
\caption{The price of American put option computed by FD and LS for different $T$ and $\alpha$ (upper panel) with corresponding running time of the algorithms (lower panel).
For small values of $T$, LS does not match the real solution properly. FD is precise and fast method for all $T$ and $\alpha$. We take $\theta=\check{\theta_{\alpha}}$.}
\end{figure}

\begin{figure}[ht]
\raggedleft
\includegraphics[scale=0.39]{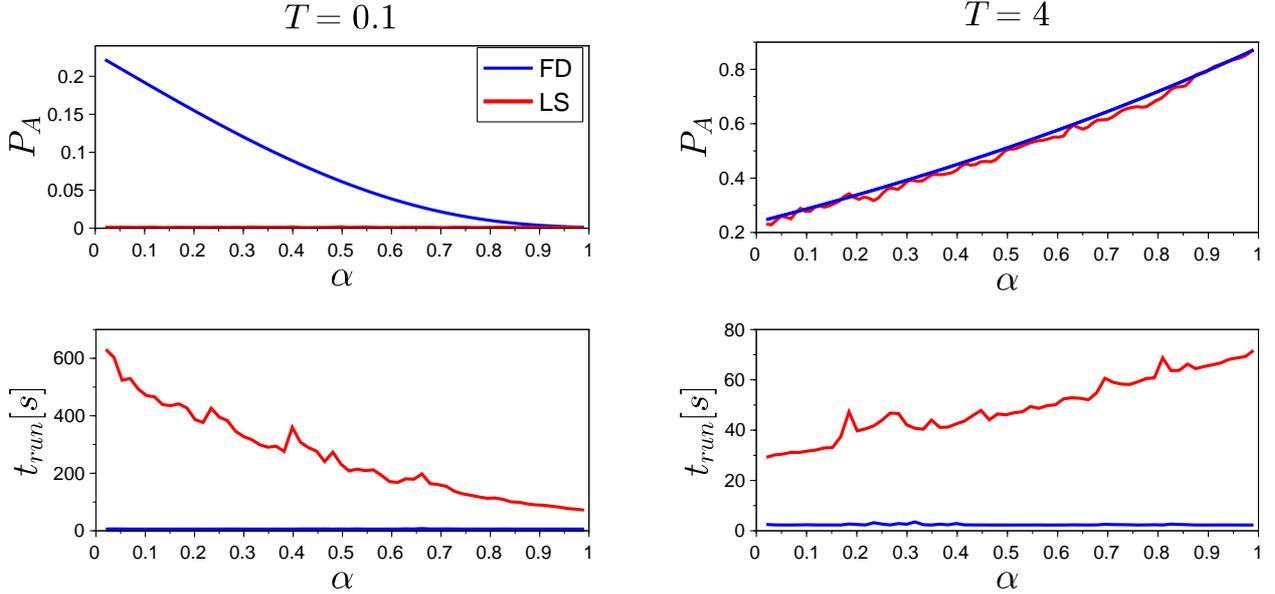}
\caption{The price of American put option computed by FD and LS for different $\alpha$ (upper panels) and corresponding running time of the algorithms (lower panels).
For small values of $T$ (left panels), LS is working visibly slower and is producing higher error as $\alpha$ is decreasing. For bigger $T$ (right panels), LS is matching the FD output. For this case, increasing $m$ and $M$ follows approaching LS to the FD output, but also increases time of computation. FD is precise and fast method for all $\alpha$ and $T$. We take $\theta=0$.}
\end{figure}

Figure $6$ presents a comparison between methods FD and LS for $3$ particular choices of $\alpha$.
We see, that as $T$ increases, the results of both methods are closer (because LS algorithm is based on the Monte Carlo, oscillations are visible).
As $\alpha$ gets higher, the LS result is closer to the FD output. Also the time of computation is visibly higher for  lower values of $T$ and $\alpha$. Note that for $T$ close to $0$ option prices manifested by different parameters $\alpha$ can significantly differ - like in the left end of the time interval $T=0,01$ (see the values computed by FD method). For each $\alpha\in(0,1)$ and considered in this example parameters, the corresponding option price is going to $0$ as $T\rightarrow 0$ but the dynamic of this convergence is not the same.
In Figure $7$ there is a comparison between FD and LS in estimating the fair price of the
American put option $P_{A}$ for ``small'' and ``big'' parameter $T$.  Also the relation between running time of both methods $t_{run}$ and $\alpha$ is provided. In both figures the FD is the reference method.
The LS method requires to generate (and save) $M$ paths of $Z_{\alpha}$ and that is memory/speed expensive. As $\alpha$ decreases, more time for generating $Z_{\alpha}(t)$ is required.
For $t<1$, $ES_{\alpha}(t)>t$, so the dynamics of $Z_{\alpha}(t)$ is "faster" than its classical equivalent ($t\in[0,T]$). Thus, the errors caused by approximating the non-Markovian process by Markovian approach can accumulate with errors caused by
ill-condition of the regression problem.
Even for "big" $T$ the LS method is running visibly longer and has lower precision than the FD. The advantage of LS is finding sample of optimal exercise times, which is not provided by the FD.
We can conclude that the presented LS algorithm could not work correctly for "small" $T$ (in particular if $T<1$). For small $\alpha$ the method is inefficient. The use of the method should be considered only
if there is a need of optimal exercise strategy. For computing the fair price of American option the FD approach is recommended.


\begin{Example}
Let us take parameters $T=4$, $Z_{0}=2$, $\sigma=0,3$, $r=0,03$, $K=2$, $H^{-}=1$, $x_{max}=100$. We focus on knock down-and-out call option. In Table \ref{table:7} and Table \ref{table:8} we compare the implicit FD method introduced in \cite{zhang} with the optimal numerical scheme (for $\theta=\check{\theta_{\alpha}}$ \cite{ja2}). Note that the implicit FD method considered in \cite{zhang} is the particular case $\theta=0$ of the weighted FD method introduced in this paper. In Table \ref{table:7} we investigate the relative error in the case $\alpha=1$ depending on $n$ and $N$. Note that for $\alpha=1$, the optimal scheme for the parameter $\theta$ $(\check{\theta_{\alpha}}=0,5)$ is called the Crank-Nicolson (C-N) method. For calculating the exact price of the option we use the analytical formula \cite{musiela}: $C^{d-o}=Z_{0}\Phi(y_{1})-Ke^{-rT}\Phi(y_{1}-\sigma\sqrt{T})-Z_{0}(H^{-}/Z_{0})^{2\lambda\sigma^{-2}+2}\Phi(y_{2})+Ke^{-rT}(H^{-}/Z_{0})^{2\lambda\sigma^{-2}}\Phi(y_{2}-\sigma\sqrt{T})$, where $\Phi$ denotes cumulative distribution function of the standard normal distribution, $\lambda=r-\sigma^{2}/2$, $y_{1}=\frac{\ln(Z_{0}/K)+(r+\sigma^{2}/2)T}{\sigma\sqrt{T}}$, $y_{2}=\frac{\ln((H^{-})^2/(KZ_{0}))+(r+\sigma^{2}/2)T}{\sigma\sqrt{T}}$. The C-N method is more efficient than its implicit equivalent. The same conclusion is true for American options (see e.g. \cite{ja}). 

\begin{table}[!ht]\footnotesize
\begin{center}
\begin{tabular}{|c c c c c c c|}
\hline
$(n,N)$&$(20,20)$ &$(40,40)$& $(100,100)$&  $(200,200)$&$(500,500)$&$(1500,1500)$\\ \hline
$\theta=0$& $1,98\%$&$1,03\%$ &   $0,39\%$  &  $0,18\%$&$0,06\%$&$0,01\%$\\			
$\theta=\check{\theta_{\alpha}}$& $0,55\%$&$0,28\%$&   $0,07\%$  &  $0,02\%$ &$0$&$0$\\			

\hline
\end{tabular}
\caption{\label{table:7}Relative error for knock down-and-out call option. The real value is $0,56$.} 
\end{center}
\end{table}

\begin{table}[!ht]\footnotesize
\begin{center}
\begin{tabular}{|c c c c c c c|}
\hline
$(n,N)$&$(20,20)$ &$(40,40)$& $(100,100)$&  $(200,200)$&$(500,500)$&$(1500,1500)$\\ \hline
$\theta=0$& $8,04\times 10^{3}$&$0,02$ &   $0,16$  &  $0,56$&$4,35$&$43,73$\\			
$\theta=\check{\theta_{\alpha}}$& $7,25\times 10^{3}$&$0,02$&   $0,15$  &  $0,56$ &$4,34$&$43,74$\\			

\hline
\end{tabular}
\caption{\label{table:7a}The average running time $[s]$ of the implicit and optimal FD method related to Table \ref{table:7}.}
\end{center}
\end{table}

In Table \ref{table:8} there are presented the relative errors (columns $6$-$9$) and the differences between the option prices calculated for implicit and optimal numerical schemes (columns $3$-$5$) for different $n,N$ and $\alpha$. The most significant difference is observed for high $\alpha$, since $\check{\theta_\alpha}$ (column $2$) is increasing function of $\alpha$. As $\alpha$ approaches $0$, the optimal scheme is reducing to the implicit method. The difference for $n=N=1500$ for each investigated value of $\alpha$ is lower than $10^{-4}$. Since the differences between both prices for $n=N=3000$ are negligible this common value we use as an exact price to compute the relative errors for the methods. We can observe that the implicit method is less accurate than the optimal FD method proposed in this paper. In the example, to compute the prices of knock down-and-out call option by implicit and "optimal" methods, we used the numerical scheme (\ref{13}) with initial-boundary conditions (\ref{18})-(\ref{19}) for $\theta=0$ and $\theta=\check{\theta_{\alpha}}$ respectively. In Table \ref{table:7a} and Table \ref{table:8a} the corresponding (mean) running times to Table \ref{table:7} and Table \ref{table:8} respectively are presented. The corresponding values to $(n,N)=(1500,1500)$ were calculated by (arithmetic) mean based on $100$ simulations, the others were computed by (arithmetic) mean based on $1000$ simulations. With increasing precision (i.e. with higher values of $n$ and $N$) the algorithm works longer. In contrast to Table \ref{table:7} and Table \ref{table:8} the significant differences between both numerical schemes were not observed.

\begin{table}[!ht]\footnotesize
\begin{center}
\begin{tabular}{|c c c c  c c c c c|}
\hline
 & \arrvline &\multicolumn{3}{c|}{$C^{d-o}(\check{\theta_{\alpha}})-C^{d-o}(0)$} & \multicolumn{2}{c}{relative error $\theta=0$}& \multicolumn{2}{c|}{relative error $\theta=\check{\theta_{\alpha}}$}\\
\hline
$\alpha$& $\check{\theta_{\alpha}}$ &$(n,N)=(20,20)$&   $(200,200)$  &  $(1500,1500)$& $(40,40)$& $(100,100)$& $(40,40)$ & $(100,100)$\\ \hline
$0,9$&$0,48$&     $0,65\%$ &   $0,07\%$  &  $0,01\%$         &$1,01\%$ &$0,39\%$& $0,36\%$&$0,12\%$\\
$0,8$&$0,46$& $0,51\%$ &   $0,05\%$  &  $0$               &  $0,91\%$&$0,35\%$& $0,36\%$&$0,13\%$\\	
$0,7$&$0,44$& $0,39\%$ &   $0,04\%$  &  $0$             &  $0,78\%$&$0,31\%$& $0,33\%$&$0,13\%$\\		
$0,6$&$0,41$& $0,29\%$ &   $0,03\%$  &  $0$             &  $0,64\%$&$0,26\%$& $0,28\%$&$0,12\%$\\
$0,5$&$0,37$& $0,20\%$ &   $0,02\%$  &  $0$             &  $0,5\%$&$0,22\%$& $0,23\%$&$0,11\%$\\
$0,4$&$0,33$& $0,13\%$ &   $0,01\%$  &  $0$           &  $0,36\%$&$0,18\%$& $0,17\%$&$0,11\%$\\
$0,3$&$0,27$& $0,07\%$ &   $0$  &  $0$         &  $0,22\%$&$0,15\%$& $0,11\%$&$0,1\%$\\			
\hline
\end{tabular}
\caption{\label{table:8}The differences and relative errors for the option prices calculated by optimal and implicit numerical schemes for different $n,N$ and $\alpha$.}
\end{center}
\end{table}

\begin{table}[!ht]\footnotesize
\begin{center}
\begin{tabular}{|c c c c  c c c c c c c|}
\hline
  $\alpha$ & \multicolumn{5}{|c|}{$\theta=0$}& \multicolumn{5}{c|}{$\theta=\check{\theta_{\alpha}}$}\\
\hline
&$(n,N)=(20,20)$&   $(40,40)$  &  $(100,100)$& $(200,200)$& $(1500,1500)$& $(20,20)$&   $(40,40)$  &  $(100,100)$& $(200,200)$& $(1500,1500)$\\ \hline
$0,9$& $0,01$ &   $0,04$  &  $0,24$         &$0,51$ &$45,64$		& $0,01$&$0,04$ &  $0,24$&$0,51$ &$45,84$\\
$0,8$& $0,02$ &   $0,05$  &  $0,28$         &$0,51$&$47,53$		& $0,02$&$0,05$ &  $0,28$ &$0,51$&$46,77$\\	
$0,7$& $0,02$ &   $0,03$  &  $0,15$         &$0,54$&$43,34$			& $0,02$&$0,03$ &  $0,15$ &$0,53$&$42,98$\\		
$0,6$& $0,02$ &   $0,04$  &  $0,14$         &$0,54$&$43,61$			& $0,02$&$0,04$ &  $0,14$ &$0,54$ &$43,55$ \\
$0,5$& $0,02$ &   $0,04$  &  $0,14$         &$0,74$&$43,49$			& $0,02$&$0,04$ &  $0,14$ &$0,74$&$43,65$\\
$0,4$& $0,02$ &   $0,03$  &  $0,14$         &$0,72$&$45,2$			& $0,02$&$0,03$ & $0,14$ &$0,72$&$45,65$\\
$0,3$& $0,02$ &   $0,04$  &  $0,14$			&$0,52$&$45,42$			& $0,02$&$0,04$ &  $0,14$ &$0,52$&$46,05$\\			
\hline
\end{tabular}
\caption{\label{table:8a}The average running time $[s]$ of the implicit and optimal FD method related to Table \ref{table:8}.}
\end{center}
\end{table}

\end{Example}

\section{Summary}
In this paper:
\begin{itemize}
\item[--] We have derived the system describing the fair price of American put option in subdiffusive B-S model.
\item[--] We have introduced the weighted numerical scheme for this system.
\item[--] We have shown how to modify previous results for valuing wide range of barrier options in frame of the same model.
\item[--] We have given the formula for the optimal choice of weighting parameter $\theta$ in dependence of subdiffusion parameter $\alpha$. 
\item[--] We have applied the Longstaff-Schwartz numerical method for the subdiffusive B-S model. This method is worse than the FD method in terms of speed and precision of computation. Moreover for small values of $T$ this method is unstable so can not be used in many different cases. 
\item[--] We have presented some numerical examples to illustrate introduced theory.
\end{itemize}

The numerical techniques presented in this paper can successfully be repeated for other time fractional diffusion models with moving boundaries problems.
\section*{Acknowledgements}
This research was partially supported by NCN Sonata Bis 9 grant nr 2019/34/E/ST1/00360.
\appendix
\section{Longstaff-Schwartz Algorithm}
The method uses a dynamic programming to find the optimal stopping time and Monte
Carlo to approximate the fair price of an option. We start assuming the exercise time $\tau$ is equal $T$. Going
backwards to $0$, we replace $\tau$ by each moment we find where is better to exercise.
Let us denote $V(Z(t),t)$ as the fair price of an American option with the payoff function $f(Z(t))$ and underlying asset $Z(t)$.
It is easy to conclude that
\[V(Z_{0},0)=E\left(e^{-r\tau}f(Z(\tau))\right).\]
Let us divide the interval $[0,T]$ into $m$ subintervals (of the same length) using the grid $[t_{0}=0,t_{1},\ldots,t_{m}=T]$, moreover we introduce
\begin{equation}
 H(Z(t_{i}))=E\left(e^{-r\left(\tau_{i}-t\right)}f(Z(\tau_{i}))|Z(t_{i})\right),\label{39}
\end{equation}
where $\tau_{i}$ is the optimal exercise moment in $\{t_{i+1},\ldots,t_{m-1},t_{m}\}$, $i=0,1\ldots,m-2$.
The interpretation of the function $H(Z(t_{i}))$ is the expected profit from keeping the option up to time $t_{i}$.
For each trajectory we will proceed using the following algorithm:

\begin{algorithm}
\caption{}
\begin{algorithmic}[1]
\State $\tau=t_{m}$, $V=f(Z(\tau))$
\For{$t$ from $t_{m-1}$ to $t_{1}$}
\State $V\leftarrow e^{-r\Delta t}V$
\If{$H(Z(t))<f(Z(t))$}
\State	$\tau\leftarrow t$
\State	$V\leftarrow f(Z(\tau))$
\EndIf
\EndFor
\State $V\leftarrow e^{-r\Delta t}V$
\State \textbf{Return} $V$.
\end{algorithmic}
\end{algorithm}

In other words at each grid point from $\{t_{1},\ldots,t_{m-1}\}$ we compare profit from keeping and exercise the option. Then we decide is it more profitable to exercise the option or keep it on. In the algorithm above $V=V(Z(0),0)$.
The key question is how to estimate values $H(Z(t_{i}))$ for $i=1,\ldots,m$. To do so, Longstaff and Schwartz  proposed to use least
squares regression.  This can be done since the conditional expectation is an
element of $L^{2}$ space, so it can be represented using its infinite countable orthonormal basis. To proceed with the computations, the finite set of $l$ such
basis elements should be chosen. For the simulations we choose 3 first elements of Laguerre polynomials
$L_{0}(x)=1$, $L_{1}(x)=1-x$, $L_{2}(x)=1/2\left(2-4x-x^{2}\right)$. Note that the early exercise at $t$ can be profitable only if $f(Z(t))>0$, i.e. if option is in the money.
The whole LS algorithm for the subdiffusive case will look as follows:

\begin{algorithm}
\caption{}
\begin{algorithmic}[1]
\State Generate $Z_{j}(t_{i})$ \cite{MM} for $j=1\ldots,M$, $i=1\ldots,m$
\State $\tau=[t_{m},t_{m},\ldots,t_{m}]$, $V=f(Z(\tau))$
\For{$t$ from $t_{m-1}$ to $t_{1}$}
\State Find in the money trajectories i.e. $w=\{j_{1},\ldots,j_{R}\}$ s.t. $f(Z_{k}(t))>0$ for $k\in w$
\State Put $Z_{w}\leftarrow[Z_{j_{1}},\ldots, Z_{j_{R}}]$, $V_{w}\leftarrow[V_{j_{1}},\ldots, V_{j_{R}}]$
\State Find regression coefficients $\beta_{0},\ldots,\beta_{l}$ such that $\displaystyle\sum_{i=0}^{l}\beta_{i} L_{i}(Z_{w})=e^{-r\Delta t}V_{w}$,
\State For $k\in w$
\If{$\displaystyle\sum_{i=0}^{l}\beta_{i} L_{i}(Z_{k})<f(Z_{k}(t))$}
\State	$\tau_{k}\leftarrow t$
\State	$V_{k}\leftarrow f(Z(\tau_{k}))$
\EndIf
\For{$i\in\{1\ldots,M\}\setminus w$}
\State $V_{i}\leftarrow e^{-r\Delta t}V_{i}$
\EndFor
\EndFor
\State $Price\displaystyle \leftarrow \sum_{i=1}^{M}\frac{e^{-r\Delta t}V_{i}}{M}$
\State \textbf{Return} $Price$.
\end{algorithmic}
\end{algorithm}


\begin{thebibliography}{100}

\bibitem{com2}
Hazhir Aliahmadi, Mahsan Tavakoli-Kakhki, and Hamid Khaloozadeh.
\newblock Option pricing under finite moment log stable process in a regulated
  market: a generalized fractional path integral formulation and monte carlo
  based simulation.
\newblock {\em Communications in Nonlinear Science and Numerical Simulation},
  page 105345, 2020.

\bibitem{Dodan2}
Ghada Alobaidi, Roland Mallier, and A~Stanley~Deakin.
\newblock Laplace transforms and installment options.
\newblock {\em Mathematical Models and Methods in Applied Sciences},
  14(08):1167--1189, 2004.

\bibitem{nd2}
Luca~Vincenzo Ballestra.
\newblock {Fast} and accurate calculation of {American} option prices.
\newblock {\em Decisions in Economics and Finance}, 41(2):399--426, 2018.

\bibitem{fast1}
Luca~Vincenzo Ballestra and Liliana Cecere.
\newblock A fast numerical method to price {American} options under the bates
  model.
\newblock {\em Computers \& Mathematics with Applications}, 72(5):1305--1319,
  2016.

\bibitem{bor}
Szymon Borak, Adam Misiorek, and Rafa{\l} Weron.
\newblock Models for heavy-tailed asset returns.
\newblock In {\em Statistical tools for finance and insurance}, pages 21--55.
  Springer, 2011.

\bibitem{fast3}
Michael~J Brennan and Eduardo~S Schwartz.
\newblock The valuation of {American} put options.
\newblock {\em The Journal of Finance}, 32(2):449--462, 1977.

\bibitem{Bar2}
Paul Brockman and Harry~J Turtle.
\newblock A barrier option framework for corporate security valuation.
\newblock {\em Journal of Financial Economics}, 67(3):511--529, 2003.


\bibitem{cen1}
Zhongdi Cen and Wenting Chen. \newblock A HODIE finite difference scheme for pricing American options.
\newblock{\em Advances in Difference Equations} 2019.1: 67, 2019.
\bibitem{cen2}
Zhongdi Cen and Anbo Le. 
\newblock A robust finite difference scheme for pricing American put options with Singularity-Separating method.
\newblock{\em Numerical Algorithms}, 53.4: 497-510, 2010.


\bibitem{nd3}
Chuang-Chang Chang, San-Lin Chung, and Richard~C Stapleton.
\newblock {Richardson} extrapolation techniques for the pricing of
  {American-style} options.
\newblock {\em Journal of Futures Markets: Futures, Options, and Other
  Derivative Products}, 27(8):791--817, 2007.

\bibitem{Bar1}
Bernd Engelmann, Matthias~R Fengler, Morten Nalholm, and Peter Schwendner.
\newblock Static versus dynamic hedges: an empirical comparison for barrier
  options.
\newblock {\em Review of Derivatives Research}, 9(3):239--264, 2006.

\bibitem{AW13}
Eugene~F Fama.
\newblock Risk, return and equilibrium: some clarifying comments.
\newblock {\em The Journal of Finance}, 23(1):29--40, 1968.

\bibitem{fia}
FIA.
\newblock {Global} {Futures} and {Options} {Trading} {Reaches} {Record} {Level}
  in 2019, 2019.

\bibitem{for1}
Peter~A Forsyth and Kenneth~R Vetzal.
\newblock {Quadratic convergence for valuing American} options using a penalty
  method.
\newblock {\em SIAM Journal on Scientific Computing}, 23(6):2095--2122, 2002.

\bibitem{AW16}
V~Yu Gonchar, AV~Chechkin, EL~Sorokovoi, VV~Chechkin, LI~Grigor’eva, and
  ED~Volkov.
\newblock Stable {L{\'e}vy} distributions of the density and potential
  fluctuations in the edge plasma of the u-3m torsatron.
\newblock {\em Plasma Physics Reports}, 29(5):380--390, 2003.

\bibitem{smoothing}
Steve Heston and Zhou Guofu. 
\newblock On the rate of convergence of discrete‐time contingent claims.
\newblock {\em Mathematical Finance}, 10.1: 53-75, 2000.

\bibitem{Hull}
John Hull and Alan White.
\newblock The pricing of options on assets with stochastic volatilities.
\newblock {\em The journal of finance}, 42(2):281--300, 1987.


\bibitem{istnienie}
Patrick Jaillet, Damien Lamberton, and Bernard Lapeyre.
\newblock Variational inequalities and the pricing of American options.
\newblock {\em Acta Applicandae Mathematica}, 21.3 (1990): 263-289.


\bibitem{AW15}
Aleksander Janicki and Aleksander Weron.
\newblock Can one see $\alpha$-stable variables and processes?
\newblock {\em Statistical Science}, pages 109--126, 1994.


\bibitem{corn}
Ralf Korn, Elke Korn, and Gerald Kroisandt.
\newblock {\em Monte Carlo methods and models in finance and insurance}.
\newblock CRC press, 2010.

\bibitem{ja}
Grzegorz Krzy{\.z}anowski.
\newblock Selected applications of differential equations in {Vanilla}
  {Options} valuation.
\newblock {\em Mathematica Applicanda}, 46(2), 2018.

\bibitem{ja2}
Grzegorz Krzy{\.z}anowski, Marcin Magdziarz, and \L{}ukasz P\l{}ociniczak.
\newblock A weighted finite difference method for subdiffusive
  {B}lack–{S}choles model.
\newblock {\em Computers \& Mathematics with Applications}, 80(5):653 -- 670,
  2020.

\bibitem{bdt}
Grzegorz Krzy{\.z}anowski, Ernesto Mordecki, and Andr{\'e}s Sosa.
\newblock A zero interest rate {Black-Derman-Toy} model.
\newblock {\em arXiv preprint arXiv:1908.04401}, 2019.

\bibitem{com1}
Sha Lin and Xin-Jiang He.
\newblock A regime switching fractional {Black--Scholes} model and european
  option pricing.
\newblock {\em Communications in Nonlinear Science and Numerical Simulation},
  85:105222, 2020.

\bibitem{Lin}
Yumin Lin and Chuanju Xu. \newblock Finite difference/spectral approximations for the time-fractional diffusion equation. \newblock {\em Journal of computational physics} 225.2 (2007): 1533-1552.

\bibitem{Davis2}
Hsuan-Ku Liu and Jui-Jane Chang.
\newblock A closed-form approximation for the fractional {B}lack--{S}choles
  model with transaction costs.
\newblock {\em Computers \& Mathematics with Applications}, 65(11):1719--1726,
  2013.

\bibitem{LSswap}
Francis~A Longstaff, Pedro Santa-Clara, and Eduardo~S Schwartz.
\newblock Throwing away a billion dollars: The cost of suboptimal exercise
  strategies in the swaptions market.
\newblock {\em Journal of Financial Economics}, 62(1):39--66, 2001.

\bibitem{LS}
Francis~A Longstaff and Eduardo~S Schwartz.
\newblock Valuing {American} options by simulation: a simple least-squares
  approach.
\newblock {\em The review of financial studies}, 14(1):113--147, 2001.

\bibitem{madi1}
Sofiane Madi, Mohamed~Cherif Bouras, Mohamed Haiour, and Andreas Stahel.
\newblock {Pricing of American options, using the Brennan--Schwartz} algorithm
  based on finite elements.
\newblock {\em Applied Mathematics and Computation}, 339:846--852, 2018.

\bibitem{MM}
Marcin Magdziarz.
\newblock {Black-Scholes} formula in subdiffusive regime.
\newblock {\em Journal of Statistical Physics}, 136(3):553--564, 2009.

\bibitem{Gajda}
Marcin Magdziarz and Janusz Gajda.
\newblock Anomalous dynamics of {Black-Scholes} model time-changed by inverse
  subordinators.
\newblock {\em Acta Physica Polonica B}, 43(5), 2012.

\bibitem{AW12}
Benoit~B Mandelbrot.
\newblock The variation of certain speculative prices.
\newblock In {\em Fractals and scaling in finance}, pages 371--418. Springer,
  1997.

\bibitem{element1}
SAJID Memon.
\newblock {Finite element method for American} option pricing: a penalty
  approach.
\newblock {\em International Journal of Numerical Analysis and Modelling,
  Series B: Computing and Information}, 3(3):345--370, 2012.

\bibitem{Merton}
Robert~C Merton.
\newblock On the pricing of corporate debt: The risk structure of interest
  rates.
\newblock {\em The Journal of finance}, 29(2):449--470, 1974.

\bibitem{AW17}
T~Mizuuchi, VV~Chechkin, K~Ohashi, EL~Sorokovoy, AV~Chechkin, V~Yu Gonchar,
  K~Takahashi, S~Kobayashi, K~Nagasaki, H~Okada, et~al.
\newblock Edge fluctuation studies in heliotron j.
\newblock {\em Journal of nuclear materials}, 337:332--336, 2005.


\bibitem{niestabLS}
Oleksii Mostovyi.
\newblock On the stability the least squares monte carlo.
\newblock {\em Optimization Letters}, 7(2):259--265, 2013.

\bibitem{musiela}
Marek Musiela and Marek Rutkowski.
\newblock {Martingale} methods in financial modelling, 2005, 2005.

\bibitem{nd1}
JC~Ndogmo and DB~Ntwiga.
\newblock {High Order Accurate Implicit Methods for Barrier Option Pricing}.
\newblock {\em Applied Mathematics and Computation}, 2011.

\bibitem{for2}
Bj{\o}rn~Fredrik Nielsen, Ola Skavhaug, and Aslak Tveito.
\newblock {Penalty and front-fixing methods for the numerical solution of
  American} option problems.
\newblock {\em Journal of Computational Finance}, 5(4):69--98, 2002.

\bibitem{orzel}
Sebastian Orze{\l} and Aleksander Weron.
\newblock {Calibration} of the subdiffusive {Black-Scholes} model.
\newblock {\em Acta Phys. Pol. B}, 41(5):1051--1059, 2010.

\bibitem{AW14}
Svetlozar Rachev and Stefan Mittnik.
\newblock Stable {Paretian} models in finance.
\newblock {\em Vol. 7. Wiley}
\newblock 2000.

\bibitem{fast2}
Jamal~Amani Rad, Kourosh Parand, and Luca~Vincenzo Ballestra.
\newblock {Pricing European and American} options by radial basis point
  interpolation.
\newblock {\em Applied Mathematics and Computation}, 251:363--377, 2015.

\bibitem{bin11}
LCG Rogers and EJ~Stapleton.
\newblock Fast accurate binomial pricing.
\newblock {\em Finance and Stochastics}, 2(1):3--17, 1997.

\bibitem{bin1}
Mark Rubinstein.
\newblock {On} the relation between binomial and trinomial option pricing
  models.
\newblock {\em The Journal of Derivatives}, 8(2):47--50, 2000.

\bibitem{sato}
Ken-iti Sato, Sato Ken-Iti, and A~Katok.
\newblock {\em L{\'e}vy processes and infinitely divisible distributions}.
\newblock Cambridge university press, 1999.

\bibitem{market}
Aleksandra Stankovska.
\newblock Global derivatives market.
\newblock {\em SEEU Review}, 12, 01 2016.

\bibitem{AW18}
Bart~W Stuck and B~Kleiner.
\newblock A statistical analysis of telephone noise.
\newblock {\em Bell System Technical Journal}, 53(7):1263--1320, 1974.

\bibitem{com3}
Dina Tavares, Ricardo Almeida, and Delfim~FM Torres.
\newblock Caputo derivatives of fractional variable order: numerical
  approximations.
\newblock {\em Communications in Nonlinear Science and Numerical Simulation},
  35:69--87, 2016.

\bibitem{weron}
Aleksander Weron and Rafa{\l} Weron.
\newblock In{\.z}ynieria finansowa (financial engineering).
\newblock {\em Wydawnictwo Naukowo-Techniczne, Warszawa}, 1998.

\bibitem{fast4}
Paul Wilmott.
\newblock {\em {Derivatives: The} theory and practice of financial
  engineering}.
\newblock John Wiley \& Son Limited, 1998.

\bibitem{zhang}
Hongmei Zhang, et al. \newblock Numerical solution of the time fractional Black–Scholes model governing European options. \newblock {\em Computers \& Mathematics with Applications}, 71.9 (2016): 1772-1783.

\bibitem{compact}
Jichao Zhao, Matt Davison, and Robert~M Corless.
\newblock {Compact} finite difference method for {American} option pricing.
\newblock {\em Journal of Computational and Applied Mathematics},
  206(1):306--321, 2007.
\end{thebibliography}

\end{document}